\documentclass[preprint,authoryear, 12pt]{elsarticle}

\usepackage[utf8]{inputenc}
\usepackage{amsmath}
\usepackage{lineno}
\usepackage{amssymb}
\usepackage{xcolor}
\usepackage{cancel}
\usepackage{graphicx}
\usepackage{outlines}
\usepackage[authoryear]{natbib}
\usepackage{bm}
\usepackage{setspace}

\journal{Ocean Modelling}

\begin{document}
\begin{frontmatter}
\doublespacing

\title{Assimilation of distributed ocean wave sensors}
\author[1]{Smit P.B.\corref{cor1}}
\author[1,2]{Houghton I.A.}
\author[1]{Jordanova K. }
\author[1]{Portwood T.}
\author[1]{Shapiro E.}
\author[1]{Clark  D.}
\author[1]{Sosa M.}
\author[1]{Janssen T.T.}

\address[1]{Sofar Ocean Technologies, San Francisco, California, USA}
\address[2]{The Data Institute, University of San Francisco, San Francisco, California, USA}
\cortext[cor1]{Corresponding author, pieter@sofarocean.com}

\begin{abstract}
In-situ ocean wave observations are critical to improve model skill and validate remote sensing wave measurements. Historically, such observations are extremely sparse due to the large costs and complexity of traditional wave buoys and sensors. In this work, we present a recently deployed network of free-drifting satellite-connected surface weather buoys that provide long-dwell coverage of surface weather in the northern Pacific Ocean basin. To evaluate the leading-order improvements to model forecast skill using this distributed sensor network, we implement a widely-used data assimilation technique and compare forecast skill to the same model without data assimilation. Even with a basic assimilation strategy as used here, we find remarkable improvements to forecast accuracy from the incorporation of wave buoy observations, with a 27$\%$ reduction in root-mean-square error in significant waveheights overall. For an extreme event, where forecast accuracy is particularly relevant, we observe considerable improvements in both arrival time and magnitude of the swell on the order of 6 hours and 1 m, respectively. Our results show that distributed ocean networks can meaningfully improve model skill, at extremely low cost. Refinements to the assimilation strategy are straightforward to achieve and will result in immediate further modelling gains.
\end{abstract}


\end{frontmatter}

\doublespacing

\section{Introduction}
The dynamics of wind-driven waves on the surface of the ocean affect upper ocean circulation, transport and mixing \citep[e.g.,][]{Xu1994Wave-Depth,McWILLIAMS2004AnWaters}, air-sea interaction \citep[e.g.,][]{Sullivan2010DynamicsWaves, Cavaleri2012WindSystem}, shelf exchange \citep[e.g.,][]{Lentz2008ObservationsShelf}, and the dynamics of coastal areas  \citep[e.g., ][]{LonguetHiggins.M1970,Battjes1974ComputationWaves,MacMahan.J2006}. Moreover, the ability to accurately forecast ocean waves is critical for safety at sea, coastal protection and recreation, and planning of offshore operations. Consequently, the societal and economical impacts of accurate ocean wave prediction is of similar importance to our ability to predict wind over the ocean. However, the lack of open ocean long-dwell sensor networks remains a critical bottleneck on the improvement of current operational wave models, either for model validation, calibration, or augmentation through data assimilation. 

Data assimilation (DA) is widely deployed in operational atmospheric weather forecasting systems through variational techniques \citep{Bannister2017AAssimilation}. The success of DA systems in atmospheric modelling is made possible by the considerable amount of daily observational data from land-based weather stations, weather balloons, satellite remote sensing, and other observations. Data assimilation in atmospheric forecasts indirectly also improves ocean models through improved estimates of surface stresses and fluxes. For wave models specifically, improvements in the surface wind and pressure field will generally also result in higher skill in the wave forecasts (if the model is suitably tuned to the wind field). This effect can be strong due to the duality of the wave forecasting problem. Direct improvement of wave forecasts with DA is also readily possible, but requires an extensive distributed sensor network that provides long-dwell and high-fidelity wave data to be effective.

Sparsity of high-fidelity real-time wave data is a core issue for wave assimilation systems. Buoy networks provide highly accurate estimates of wave field statistics, but these networks are sparse due to the costs of deployment and maintenance. Additionally, buoys are usually deployed near the coast on the continental shelf where they have limited value for DA forecast systems. Satellite remote sensing of surface waves has been available increasingly since the 1980s \citep[altimeter, SAR, see e.g.][]{Vesecky1981RemoteImages,Ribal201933Observations}, but these estimates carry considerable uncertainty and their spatio-temporal sampling characteristics (short-dwell) limit their effectiveness for operational DA systems \citep{Lionello1992AssimilationModel,Voorrips1997AssimilationModel,Breivik1994Model,Wittman2005AssimilationIII}.

More fundamentally, a considerable part of the ocean wave field is directly forced by surface winds and near-surface pressure fluctuations in the atmosphere (i.e. sea components) \citep{Komen1994}. As a result, the forecasting problem for wind-forced conditions is mostly a forced problem rather than an initial value problem. In turn, if the wave field is corrected without also correcting the atmospheric forcing field to match, the wave field will rapidly return to the state dictated by the forcing on time scales of 1-2 days \citep{Lionello1992AssimilationModel,Wittman2005AssimilationIII}, undoing the corrections provided by DA. In contrast, for wave components decoupled from the wind (i.e. swell components) the modelling problem is mostly an initial value problem and DA corrections to these components retain their memory much longer. 

To overcome some of these difficulties, advanced ocean wave DA systems have been developed which provide partitioning of sea and swell components \citep{Hanson2001AutomatedSpectra,Portilla-Yandun2016OnSystems}, feedback to atmospheric forcing \citep{Voorrips1997AssimilationModel}, and various variational adjoint and adjoint-free systems \citep{Veeramony.J2010,Orzech.M2013,Orzech2016RecentSWAN,Panteleev2015Adjoint-freeModel}. The variational approaches, such as 3DVar and 4DVar (\citep{Bannister2017AAssimilation}, assume the modelling error is primarily driven by errors in model input (e.g. initial conditions or forcing) and attempt to optimize these fields such that the resulting model evolution minimizes some cost function (typically a weighted error of the model with regard to the observations). The advantage is that the resulting model predictions obey the model dynamics, and forecasts tend to be smooth within the 6-hourly variational interval (although discontinuities still occur in the transition between forecasts). However, variational techniques require an iterative solver, with the computational effort of each iteration comparable to the complete forward model computational effort. Further, for nonlinear wave models with complex non-local operators in spectral space, the effort in developing and maintaining an adjoint model is non-trivial \citep{Orzech.M2013}. 

Although these approaches are promising and illustrate the potential of DA strategies for ocean waves, they have been mostly ineffective, both due to lack of efficiency, and limited availability of usable data. As a result, to date, operational wave models either do not assimilate any data, or use a lower-cost sequential method. In contrast to variational methods, sequential methods adapt the model state at a given time to better fit the observed data and subsequently use the updated field as the starting point for a new forecast period \citep{Lionello1992AssimilationModel,Voorrips1997AssimilationModel,Wittman2005AssimilationIII}. Sequential methods do not constrain the modified field to ensure the solution fits (model) physics, and consequently smoothness in time of the forecast is not guaranteed.

In the current work, we present results from a large network of open ocean wave sensors deployed in the Pacific Ocean and evaluate our ability to improve forecast skill for short-range forecasts (1-2 days). Rather than focus on advanced assimilation strategies, this work aims to assess the first-order forecasting improvements possible when sufficient data is available. To that end, we drive a simple, sequential Optimized Interpolation assimilation strategy \citep{Lionello1992AssimilationModel,Voorrips1997AssimilationModel,Wittman2005AssimilationIII} with the distributed wave buoy network. We show that by creating a step function improvement in data density in open ocean regions, data assimilation can give immediate improvement in operational wave forecast systems.

In section 2 we describe the dynamic sensor array in the Pacific, the wave model is discussed in section 3, data assimilation strategies are detailed in section 4, and the models and observations compared are described in section 5. After presenting core results in section 6, we highlight the implications of this study in section 7, and discuss how the availability of large distributed networks, combined with advances in DA strategies can effectively enable considerable improvements to operational wave forecasting.

\section{Distributed Sensor Network in the Northern Pacific}
\begin{figure}[h]
\centering
\includegraphics[width=\linewidth]{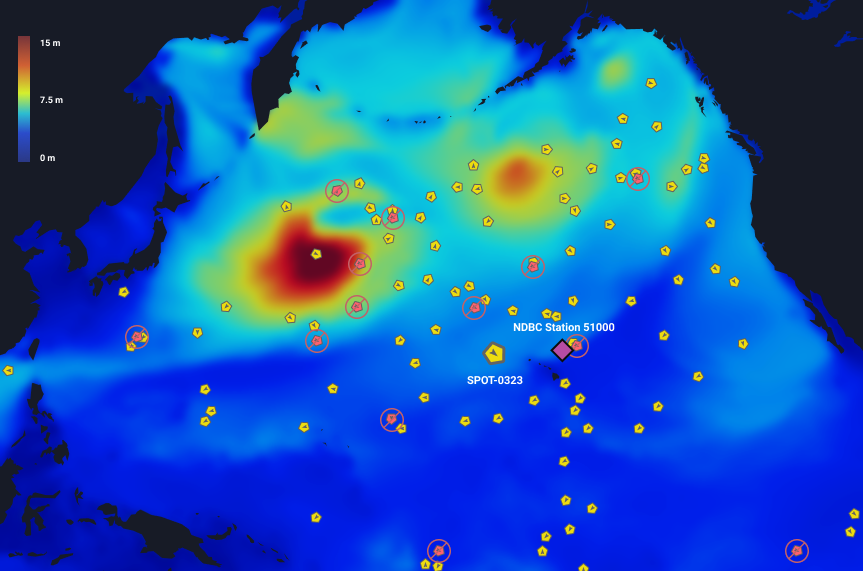}
\caption{A snapshot of the distributed sensor network on December 25$^{th}$, 2019. Spotter wave buoy locations are indicated, with yellow markers for buoys used in the re- and operational analysis and red crossed-through markers for the buoys not used in the re-analysis, but instead used for verification. The red buoys are included in the operational analysis, which is then used as the initial condition for forecast runs.}
\label{fig:dashboard}
\end{figure}

For increased wave data availability in the Northern Pacific, approximately 129 free-drifting directional wave buoys were deployed in the Pacific Ocean (as of January 1, 2020). Deployments were primarily focused on the Northern Hemisphere of the basin, where approximately 96 units were deployed. Initial deployments started in December 2018 and continued throughout 2019 with an average of 10 units coming online each month. Ongoing deployments typically focus on filling gaps in the coverage. Coverage gaps develop mostly as a consequence of the free-drifting nature of the buoys, which results in a dynamic topology of the network (figure \ref{fig:dashboard}). 

All deployed units are Sofar Spotters, an ocean wave-wind-current sensor that integrates a fast-sampling (2.5\,Hz), high-fidelity motion sensing package, onboard analysis and processing for directional wave spectra and surface drift. The Spotter buoy is compact (38 cm diameter), lightweight (5.4 kg) and completely solar-powered, which enables sustained operation without battery replacement. Units are ballasted with a \,60 cm, 2.1\,kg stainless-steel ballast chain to enhance drag and reduce wind induced drift. Spotter reproduces surface motions accurately within a frequency band from 0.03\,Hz up to approximately 1.0\,Hz. Accuracy in recording displacements is within 2\,cm under controlled conditions and validation tests show similar data quality as industry-standard wave buoys \citep[][]{Raghukumar2019PerformanceBuoy}.

The buoy samples three-dimensional displacements and estimates the vertical and horizontal displacement (cross-)spectra over an approximately 1800\,s period based on windowed Fourier estimates (window length 102.4\,s) with 50\% overlap, resulting in 33 ensemble members. From the wave spectra, standard wave bulk parameters including significant waveheight, mean spectral period, and mean direction are calculated based on conventional definitions \citep[see, e.g.][for descriptions]{Holthuijsen.L2007}. Additionally, the high-frequency tail of the wave spectrum is used to estimate local wind magnitude and direction based on equilibrium theory \citep{Voermans2020EstimatingBuoy.}.

We consider the period between July 1$^\mathrm{st}$, 2019 and December 31$^\mathrm{st}$, 2019. While Spotters can report complete displacement spectra, we only consider available bulk parameters during this period. The majority of the units reported data on an hourly interval, where each communication packet contained bulk parameters for two half-hour windows.

\section{Operational Wave model}

The evolution of wave properties on the ocean is governed by the Action Balance equation augmented with source terms for generation, dissipation, and nonlinear distribution of wave energy \citep[e.g.][]{Komen1994,Holthuijsen.L2007}. 
Mature operational systems based on this framework are typically based on either the WaveWatch 3 model \citep[WW3 hereafter,][]{Tolman1991,theWavewatchIIIDevelopmentGroup2016User5.16} as used by the National Oceanic and Atmospheric Administration (NOAA); or the WAM model \citep[][]{Wamdi1988} as used by the European Centre for Medium-range Weather Forecasts (ECMWF). Though numerical differences exist, both models implement similar physics approximations and both have proven track records in global operational forecasting settings. As such, either model provides a suitable candidate modelling framework in the present context. Here we choose to base our wave forecasts on the latest WW3 version (version 6 at the time of writing), motivated by the community driven open-source development of the model.

\subsection{Model configuration}

Wave model performance is dominated by the physics approximations of the various source terms, with source term approximations being interdependent. Calibration is typically performed on the entire set of approximations. State-of-the-art source term configurations are typically based on \citet{Ardhuin2010SemiempiricalValidation} and \citet{Zieger2015Observation-basedWAVEWATCH}, colloquially referred to as ST4 and ST6, respectively. Bulk parameters (waveheight, mean period, etc.) perform similarly between configurations \citep[e.g.][]{Stopa2016ComparisonModels,Liu2019Observation-basedVerification}, though spectral shape is improved in ST6 \citep{Liu2019Observation-basedVerification}. However, ST4 has been used operationally over an extensive period, providing a robust framework which we can easily inter-compare with present operational global NOAA forecasts. Therefore, we choose ST4 combined with the Discrete Interaction Approximation (DIA) for nonlinear interactions as our baseline system. 

The model is forced with operational wind and ice fields from the National Center for Environmental Prediction (NCEP) Global Data-assimilation System (GDAS), when available (i.e. up to and including a nowcast). GDAS winds incorporate observations available at forecast time by means of a three-dimensional data assimilation system \citep{Kleist2009IntroductionSystem}. Forecasted winds (past the nowcast time) and ice fields are obtained from the NCEP Global Forecast System (GFS). Bathymetry is derived from the Etopo2 database \citep{NationalGeophysicalDataCenter20062-minuteV2}. The present model does not include effects of meso-scale surface currents on the waves.

Model resolution is set to half of a degree for a global domain, covering ocean basins between approximately  -77.5 and 77.5 degrees latitude. Time-stepping occurs with a global time step of 30 minutes, with sub-steps in the fractional step integration sufficiently small to ensure model stability and accuracy (specifically 450\,s for spatial advection, 900\,s for intraspectral propagation and a minimum dynamic source time step of 10\,s). Further,  sub-grid features (e.g. small island chains, atolls) are included as in \citet{Tolman2003TreatmentModels}. Frequency-direction space is discretized over the full circle with 36 directions and 36 frequencies. The directional grid has a constant spacing of 10$^\circ$, whereas the frequency grid is logarithmically distributed with growfactor 1.1 (to ensure compatibility with the DIA), starting from $f_1=0.035$\,Hz and ending at $f_3=0.98$\,Hz. We differ from default ST4 configuration values for the wind source proportionality factor $\beta_{\textrm{max}}$, setting it to $\beta_{\textrm{max}}=1.32$, which has been found to improve performance with GFS based surface winds (pers. comm. with PA Wittmann). Otherwise, calibration coefficients are kept at default values for physics parameterizations.

\section{Data Assimilation}
In order to focus on evaluation of the impact of a large distributed set of sensors on model forecasts, we implement a sequential technique referred to as optimal (or statistical) interpolation. To note, optimal interpolation (OI) corresponds to the analysis step of the Kalman Filter, making an OI-based method a natural step prior to an ensemble Kalman data assimilation framework. In the current framework, the GDAS wind forecast is used along with the DA scheme described below to advance the last known wave state to the zero-time model nowcast. This produces an analysis field that is used as the initial condition for the forecast runs driven by the GFS wind fields.

\subsection{Optimal Interpolation}
Let $\bm y^{\textrm{obs}}(t)=[y^{\textrm{obs}}_1, \ldots, y^{\textrm{obs}}_{N}]^\textrm{T}$ denote the vector containing $N$ observations of a wave variable from the Spotter network at time $t$ located in the Pacific at $\bm x^{\textrm{obs}}(t)=[x^{\textrm{obs}}_1, \ldots, x^{\textrm{obs}}_{N}]^\textrm{T}$. Further, let $\bm y^{\textrm{mod}}(t)=[y^{\textrm{mod}}_1, \ldots, y^{\textrm{mod}}_{M}]$ denote model estimates (or `prior') within model cells centered at $\bm x^{\textrm{mod}}$. Finally, let $\tilde{\bm y}^{\textrm{obs}}$ and $\tilde{\bm y}^{\textrm{mod}}$ denote the `true' physical state at observed and modelled locations, respectively. The expected model error ($\epsilon^{\textrm{mod}}$) and observation error ($\epsilon^{\textrm{obs}}$) are assumed zero mean and given by,
\begin{align}
    \epsilon^{\textrm{mod}}= \bm y^\textrm{mod}- \tilde{\bm y}^\textrm{mod} & & \epsilon^{\textrm{obs}}= \bm y^\textrm{obs}- \tilde{\bm y}^\textrm{obs}.
\end{align}
Further, the difference between model and observation at observed locations, often referred to as `the innovation' in the literature, is given by
\begin{equation}
    \bm \epsilon = \bm y^{\textrm{obs}} - \mathbf{H}\bm y^\textrm{mod} =  \epsilon^{obs}- \mathbf{H}\epsilon^{\textrm{mod}} - \epsilon^{\textrm{int}}.
\end{equation}
Here $\mathbf{H}$ is the interpolation matrix that estimates model values at observed locations through bi-linear interpolation, and the interpolation error $\epsilon^{\textrm{int}}=
\mathbf{H}\tilde{\bm y}^{\textrm{mod}} - \tilde{\bm y}^{\mathbf{obs}}
$ is assumed to be small. We can then use the innovation to obtain an improved analysis estimate of the target observable on the model grid through the analysis equation
\begin{equation}
    \label{eq:analyis_equation}
    \bm y^{\textrm{an}} = \bm y^{\textrm{mod}} + \mathbf{K}\epsilon,
\end{equation}
with $\textrm{K}$ the Kalman Gain matrix, and the analysis error as $\epsilon^{\textrm{an}}=\bm y^\textrm{an}-\tilde{\bm y}^\textrm{mod}$. The Kalman Gain matrix represents the linear weights that minimize the expected value of the squared sum of the mean analysis error (assuming errors are uncorrelated), and is given by
\begin{equation}
\mathbf{K} = \mathbf{C}^\textrm{mod} \mathbf{H}^{\textrm{T}} \left[ \mathbf{H} \mathbf{C}^{\textrm{mod}}\mathbf{H}^{\textrm{T}} + \mathbf{C}^{\textrm{obs}}   \right]^{-1}.
\end{equation}
Here $\mathbf{C}^\textrm{mod}$ and $\mathbf{C}^\textrm{obs}$ are the background error covariance matrices,
\begin{align}
    \mathbf{C}^{\textrm{mod}} &= \left\langle \epsilon^{\textrm{mod}}(\epsilon^{\textrm{mod}})^{\textrm{T}} \right\rangle, &
    \mathbf{C}^{\textrm{obs}} &= \left\langle \epsilon^{\textrm{obs}}(\epsilon^{\textrm{obs}})^{\textrm{T}} \right\rangle,
\end{align}
with $\langle \ldots \rangle$ denoting the expected value. 

\subsection{Model State Update}
The analysis field is ideally used directly to correct the model state, which in the context of third generation wave models is the frequency direction energy (or more specifically action) density spectrum $E^{\textrm{mod}}(f,\theta,\bm x,t)$ (with $f$ frequency, and $\theta$ direction). However, buoys do not directly sample the directional wave spectrum. Instead, only the frequency integrated spectrum is observable, and only partial information (through the lowest order directional Fourier components) on the directional shape is available. Estimates of the full observed directional spectrum, $E^{\textrm{obs}}$, can be obtained from buoys \citep[e.g.][among others]{Longuet-Higgins1963ObservationsBuoy,lygre.A1986}, however, fitting on directional moments (or possibly on partitioned data) directly may give more stable results.

Regardless, for the period considered here the buoy network only transmitted bulk parameters, and as such no spectral shape information was available. Only mean waveheights, period, direction and directional width were available to assimilate with model predictions. For a sea state comprised only of waves originating from the same storm it is reasonable to retain the local spectral shape and solely scale energy, adjust peak location and rotate mean direction to fit local observations and obtain the analysis spectrum \citep[][]{Portilla-Yandun2016OnSystems, Lionello1992AssimilationModel, Voorrips1997AssimilationModel,Wittman2005AssimilationIII} as
\begin{equation}
\label{eq:distribution}
    E^{\textrm{an}}(f,\theta) =\alpha  E^{\textrm{mod}}( \beta f ,\theta - \Delta ),
\end{equation}
with
\begin{align}
\alpha&=\frac{1}{\beta}\left(\frac{H_s^{\textrm{an}}}{H_s^{\textrm{mod}}}\right)^2, & \beta&= \frac{T_p^{\textrm{an}}}{T_p^{\textrm{mod}}}, & \Delta = \theta^\textrm{mod}_{\textrm{mean}} - \theta_{\textrm{mean}}^{\textrm{an}}
\end{align}
and $\Delta$ as the minimum directional difference on the circle. With model winds remaining uncorrected, as they are in this scheme, any correction to the local spectrum will quickly relax to the forced state for the wind sea. For swell, corrections are more likely to persist, and the largest impact of assimilation is expected. For swell, just correcting mean energy likely encompasses the largest improvement. For mixed seas, the bulk parameters alone do not provide sufficient information to correct the individual component systems. Because we focus primarily on waveheight, and we are mostly interested in the potential impact of data-assimilation of a large fleet of drifters (and not necessarily the best scheme) we choose the simplest possible correction, correcting the total energy but otherwise setting $\beta = 1$ and $\Delta =0$ \citep{Wittman2005AssimilationIII}. 

\subsection{Background Error Estimates}
Given the possibly time-dependent background error covariance matrices, $\mathbf{C}^\textrm{obs}$ and $\mathbf{C}^\textrm{mod}$, the Kalman Gain, $\mathbf{K}$, represents the weights that minimize the average analysis error. In practice, however, knowledge of the background errors is unavailable and the error covariances have to be estimated. To this end, we decompose the covariance in terms of error-correlation and standard deviation according to
\begin{equation}
\label{eq:crosscor}
    C_{m,n}^{\textrm{mod}} = \sigma_m^{\textrm{mod}}\rho(\bm x_m,\bm x_n)\sigma_n^{\textrm{mod}},
\end{equation}
where $\rho(\bm x_m,\bm x_n)$ is the spatial cross-correlation function that effectively prescribes the spatial memory of the system and $\bm{\sigma}^{\textrm{mod}}$ is the model error standard deviation. The spatial cross-correlation depends on local climatology and topography \citep[][]{Greenslade2004BackgroundData,Portilla-Yandun2016OnSystems}, and estimating the statistical footprint for the entire Pacific basin is non-trivial. Here, we assume that $\rho$ is isotropic and homogeneous with stationary statistics, such that $\rho(\bm x_m,\bm x_n)=\rho(|\bm x_m - \bm x_n|)$. We then parameterize the correlation matrix $\bm \rho$ as
\begin{equation}
    \rho_{m,n} = \exp\left[ - \left(\frac{D(\bm x_m,\bm x_n)}{\lambda}\right)^{p} \right],
\end{equation}
following \citet{Lionello1992AssimilationModel,Voorrips1997AssimilationModel,Greenslade2004BackgroundData}. Here $D(\bm x_m,\bm x_n)$ is the great circle distance between points, the length scale $\lambda$ determines the decorrelation scale, and the power $p$ determines the peakedness. Here we set $\lambda=300$\,km and $p=3/2$ (see Appendix A). Further, we assume a constant error-variance and set ${\sigma}^{\textrm{mod}}_m= \sigma^{\textrm{mod}}$. Based on comparison between buoys and model nowcasts (without data assimilation), we set $ \sigma^{\textrm{mod}}=0.3$\,m. The homogeneous, isotropic and stationary estimate of $\rho$ is a simplification that ignores, for instance, that higher latitudes tend to experience more severe weather and as a consequence have larger model errors. However, with the principal aim of gauging the first-order effect of data-assimilation from a vastly increased supply of data on model fidelity, further improved parameter estimates are not pursued in the present work.

Observational error is composed of instrument noise and errors in the statistical estimators of mean wave parameters. For well separated observers (distances of 10 \,km or more) these errors are unlikely to be correlated between Spotters and it is reasonable to assume that $\mathbf{C}^{\textrm{obs}}$ is diagonal \citep[similar to][]{Lionello1992AssimilationModel,Greenslade2004BackgroundData} and
\begin{equation}
    C^{\textrm{obs}}_{m,n} =  \sigma^{\textrm{obs}}_m \sigma^{\textrm{obs}}_n \delta_{m,n}.
\end{equation}
While instrument noise is constant and small (on the order of centimetres for a Spotter) the statistical sampling error of mean conditions grows with increased sea state energy, consequently depending on local conditions. Regardless, we again approximate $\sigma^{\textrm{obs}}_m=\sigma^{\textrm{obs}}$. Observational error due to statistical sampling is estimated as the variance between half hourly  observations and the corresponding lowpass filtered result (i.e. the 2-hour running mean), indicating $\sigma^{\textrm{obs}}\approx 0.1$ \,m.

With these estimates in place, we find that the Kalman Gain $\mathbf{K}$ reduces to
\begin{equation}
    \label{eq:invkalman}
    \mathbf{K} = \bm \rho \mathbf{H}^\textrm{T}\left[\mathbf{H}\bm \rho \mathbf{H}^\textrm{T} + \left(\frac{\sigma^{\textrm{obs}}}{\sigma^\textrm{mod}}\right)^2\mathbf{I} \right]^{-1},
\end{equation}
As a consequence, only the relative trust we have in model and observations matters. Because both observational and model error variances are proportional to the energy in the field, we approximate the ratio as a constant $\sigma^\textrm{obs}/\sigma^\textrm{mod}\approx 0.3$. If $\sigma^\textrm{mod}$ changes on a slow scale compared with $\lambda$, eq. \eqref{eq:invkalman} approximates the full form of eq. \eqref{eq:crosscor}, and eq. \eqref{eq:invkalman} applies even if error variances change. While this is currently a heuristic approach, future development will estimate background errors directly based on observations.
\subsection{Implementation}
To optimize performance of the OI algorithm we make use of the rapid drop-off in correlation with increasing distance between points. First, $\rho$ is only calculated if $|\bm x_m - \bm x_n|\lambda^{-1} < 3$, otherwise $\rho=0$ is assumed. Secondly, we partition the Northern Pacific basin into tiles of 30 by 30 degrees (resulting in a 4 by 6 grid of tiles). To each tile we add a region of 5 degrees on all sides that overlaps with neighbouring tiles to ensure observational data across the boundary can influence the OI result within the tile. Only the state within the tile boundaries proper is updated, under the assumption that observations outside of the 5 degree border do not influence the state within the tile. 

The tiling reduces the number of observation points per tile, thus reducing the size of the matrix to be inverted, and allows for efficient parallel processing. Further, the size of the covariance matrices, which scale with the number of model points squared, reduces considerably. With these approximations, the algorithm is efficient, can be applied without significant overhead in an operational setting, and is expected to scale well with increased buoy density as the network scales.

Once the analysis waveheight (or energy) field is estimated, model energies are corrected according to eq. \eqref{eq:distribution}. Specifically, the model state at analysis time is written to storage and then modified to fit the analysis estimates of significant waveheight. Subsequently, this estimated new model state serves as the starting state for the next analysis period or forecast. 

\section{Models and Observations Considered}
To compare model performance with and without data-assimilation we consider a re-analysis of the period from July 1$^\textrm{st}$, 2019 to January 20$^\textrm{th}$, 2020. The re-analysis is performed using hourly assimilation of data and is forced with GDAS wind and ice fields. To spin up the global wave model, the model simulations are started 14 days prior to July 1$^\textrm{st}$, initialized from a quiescent ocean. The 14 days of spinup time are otherwise not used. Effectively, this re-analysis produces the exact same results as the operational analysis would have if it were operational at the time. For verification purposes, 13 semi-randomly\footnote{Buoys were manually selected from an overview map to ensure a verification buoy was present in most sectors of the northern Pacific. However, no effort was made to otherwise influence performance at selected sites} selected Spotter buoys were excluded from the Optimal Interpolation (figure \ref{fig:dashboard}).

In parallel with the re-analysis model, a hindcast was performed with the same model setup but excluding data assimilation. This allowed for evaluation of differences between a model with and without data assimilation. Secondarily, the hindcast provided a comparison of results from the operational setup and those produced by the WaveWatch 3 global model of NCEP (NCEP-WW3 hereafter). With comparable setup and the same forecast winds, results from both models should be similar (allowing for minor discrepancies, e.g. nested grids were not considered). This comparison indicated that differences between our hindcast and those produced by NCEP-WW3 are indeed minute. Therefore, differences between the analysis and NCEP-WW3 are primarily due to data-assimilation, rather than our WW3 implementation. We compare results from the analysis model to NCEP-WW3, an entirely independent data source.

In addition to the re-analysis, which evaluates improvements to the nowcast, we also investigate the influence on model forecasts. To this end, we use results from operational forecasts initialized with operational analysis nowcasts that have been running in real-time since November 2019. For these forecasts, all network buoys are used in the analysis. Again, a model without data-assimilation was run in parallel to this forecast model to gauge the impact of the data-assimilation. Forecasts from this model are again interchangeable with NCEP-WW3 results, and we principally compare with NCEP-WW3.

\section{Results}

\begin{figure}[h]
\centering
\includegraphics[width=\linewidth]{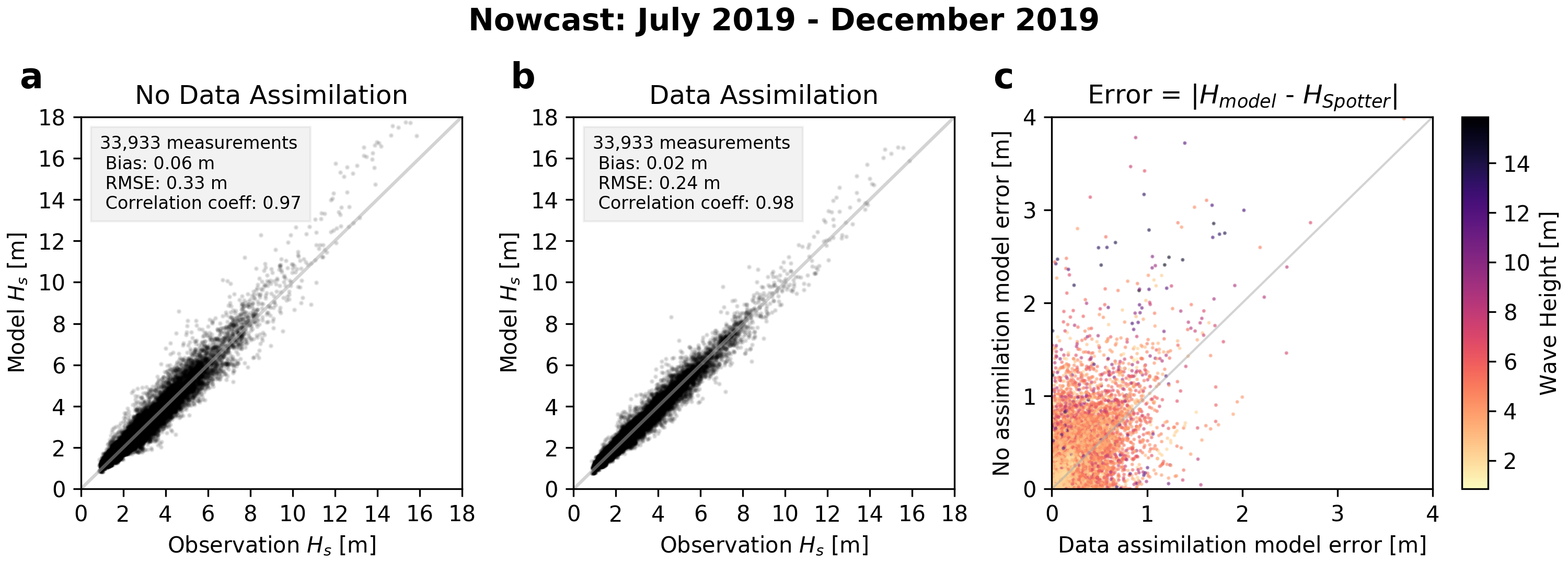}
\caption{Near real-time model predictions of significant waveheight compared to Spotter observations for July 2019 to December 2019. Only Spotter measurements not included in the data assimilation were used as observations. (a) NOAA Wavewatch 3 model without assimilation of Spotter measurements. (b) The wave model with assimilation of Spotter measurements. (c) For each waveheight measurement, model error with data assimilation versus model error without data assimilation. All points falling above the one-to-one line indicate improved model ability from assimilation of data. Measurement points are colored by observed significant waveheight.}
\label{fig:nowcast}
\end{figure}

\begin{figure}[h]
\centering
\includegraphics[width=.75\linewidth]{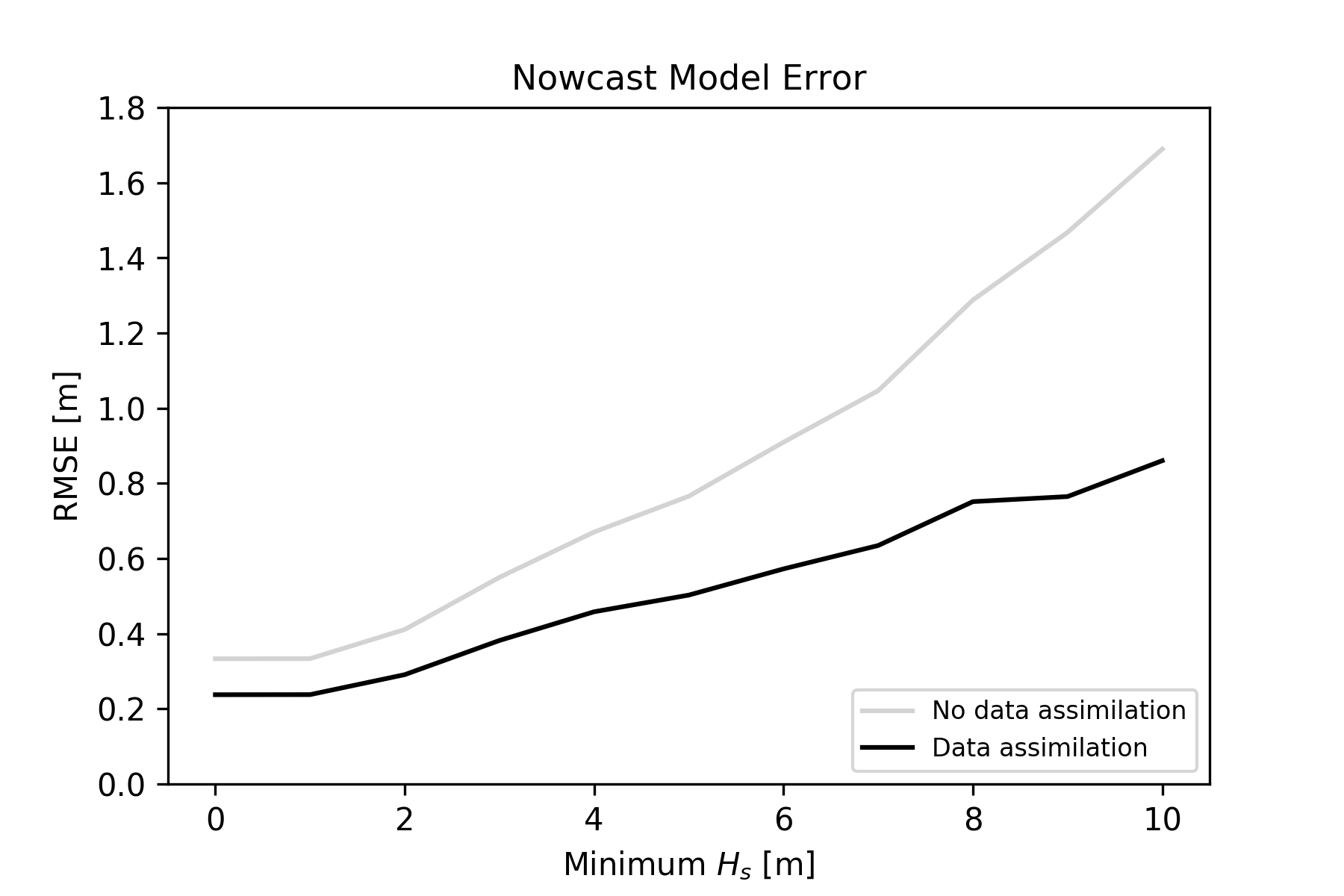}
\caption{Root-mean-square error (RMSE) for models with and without data assimilation as a function of minimum observed significant waveheight. The model with data assimilation always outperformed the model without data assimilation, with even larger improvements for predictions in sea states with large significant waveheights.}
\label{fig:nowcastRMSE}
\end{figure}

The model predictive accuracy of significant waveheight was improved by assimilation of sensor network data. Improvements in forecast accuracy were observed for both current sea state estimates and future forecasts with lead times up to four days. To estimate model improvement, the Spotter observations excluded from the data assimilation were compared with model predictions evaluated at that subset of Spotter locations. Compared to the model without data assimilation (NCEP-WW3), the data-assimilation model reduced the root-mean-square error (RMSE) by 27\% (0.33 m to 0.24 m) for all sea states (figure \ref{fig:nowcast}a-b). For both models, the waveheight error scaled with the waveheight magnitude such that larger waves led to larger errors. However, the increase in error with waveheight was smaller for the data assimilation model than the model without data assimilation (figure \ref{fig:nowcastRMSE}). That is, in energetic sea states with large waves, the data assimilation model accuracy exhibited even larger improvements, with an RMSE reduction of 35\% (0.77 m to 0.50 m) for sea states with waveheights above 5 m, and 49\% (1.69 m to 0.83 m) for waveheights above 10 m.

\begin{figure}[h]
\centering
\includegraphics[width=\linewidth]{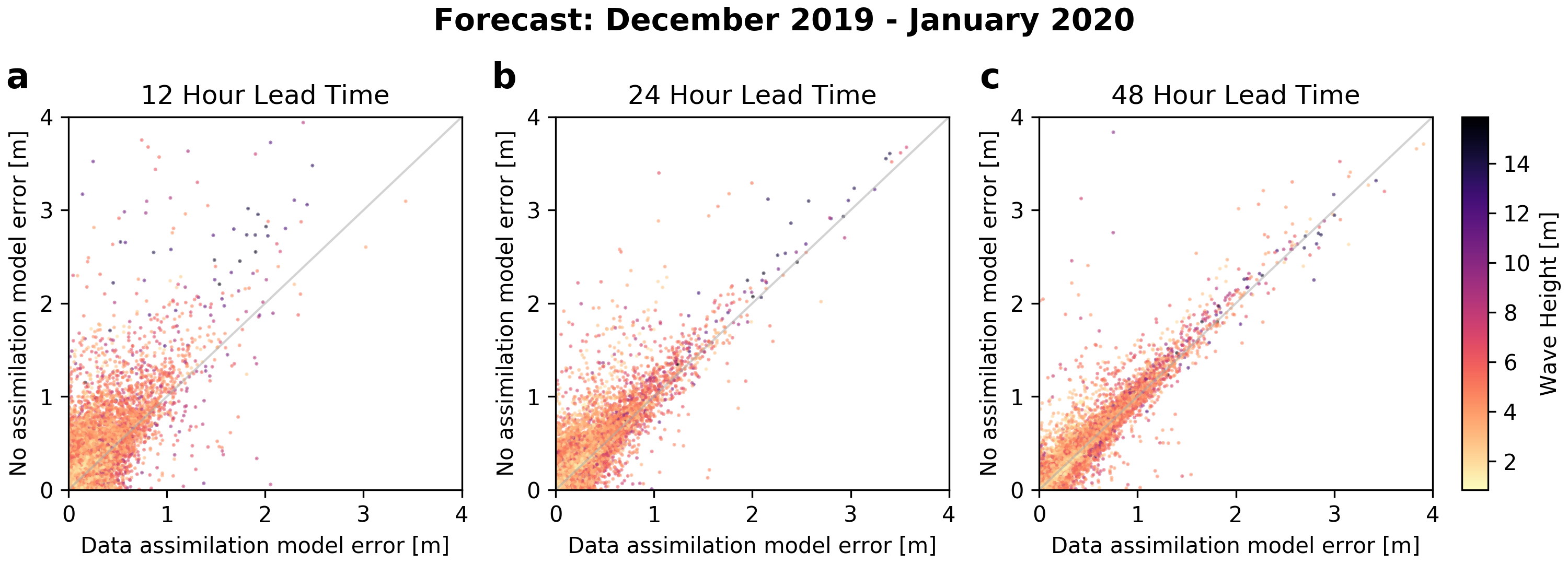}
\caption{Comparison of errors from models with and without data assimilation, similar to figure \ref{fig:nowcast}c, for (a) 12 hour forecast, (b) 24 hour forecast, and (c) 48 hour forecast. Data assimilation was implemented in future forecasts from December 2019 to January 2020. All points falling above the one-to-one line indicate improved model ability from assimilation of Spotter data.}
\label{fig:forecasts}
\end{figure}

\begin{figure}[h]
\centering
\includegraphics[width=.75\linewidth]{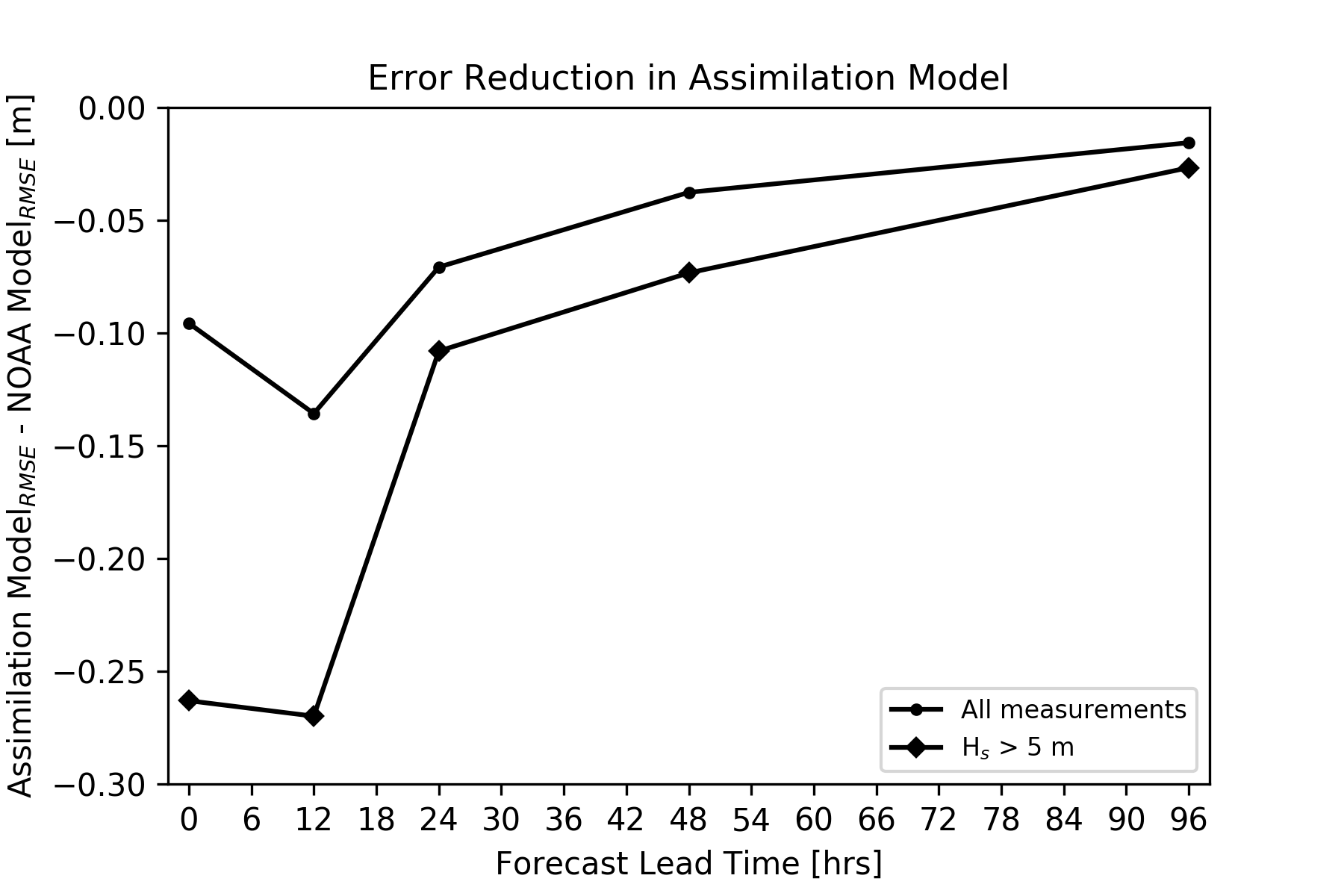}
\caption{Comparison of root-mean-square error (RMSE) of models. The difference between the RMSE with data assimilation and the RMSE without data assimilation was always negative, indicating error reduction for all forecast lead times. With all measurements considered (circles), the data assimilated model always had a smaller RMSE than that without assimilation. Sea states with large observed significant waveheight (H$_s$ $>$ 5 m, diamonds), led to even larger error reduction from data assimilation.}
\label{fig:forecastRMSE}
\end{figure}

Similarly, the model with data assimilation frequently outperformed the model without data assimilation for forecasts with lead times of 12, 24, and 48 hours (figure \ref{fig:forecasts}). The model with data assimilation had a smaller RMSE than the model without data assimilation, with the largest improvements for shorter lead times (figure \ref{fig:forecastRMSE}). The accuracy improvement from data assimilation diminished with increasing lead time, converging to negligible improvements beyond a four day lead time. Similar to the nowcast, error improvement was even larger for larger waveheights in future forecasts, with an error reduction for a lead time of 12 hours of 0.14 m for all waves and 0.27 m for waves over 5 m.  

With regard to other wave parameters (mean direction, mean period, directional spread). No significant differences are found for either nowcasted or forecasted predictions (not shown). To note, even though only mean energy (and not distribution) are corrected, predicted spectral distributions down-wave of buoys can be affected. The energy in components originating down-wave of assimilation points is modified, whereas spectral components originating elsewhere remain unchanged, thus affecting the relative contribution of each.
However, this does not appear to significantly alter forecast skill.

\subsection{December 2019 Storm Swell}

\begin{figure}[h]
\centering
\includegraphics[width=\linewidth,page=2]{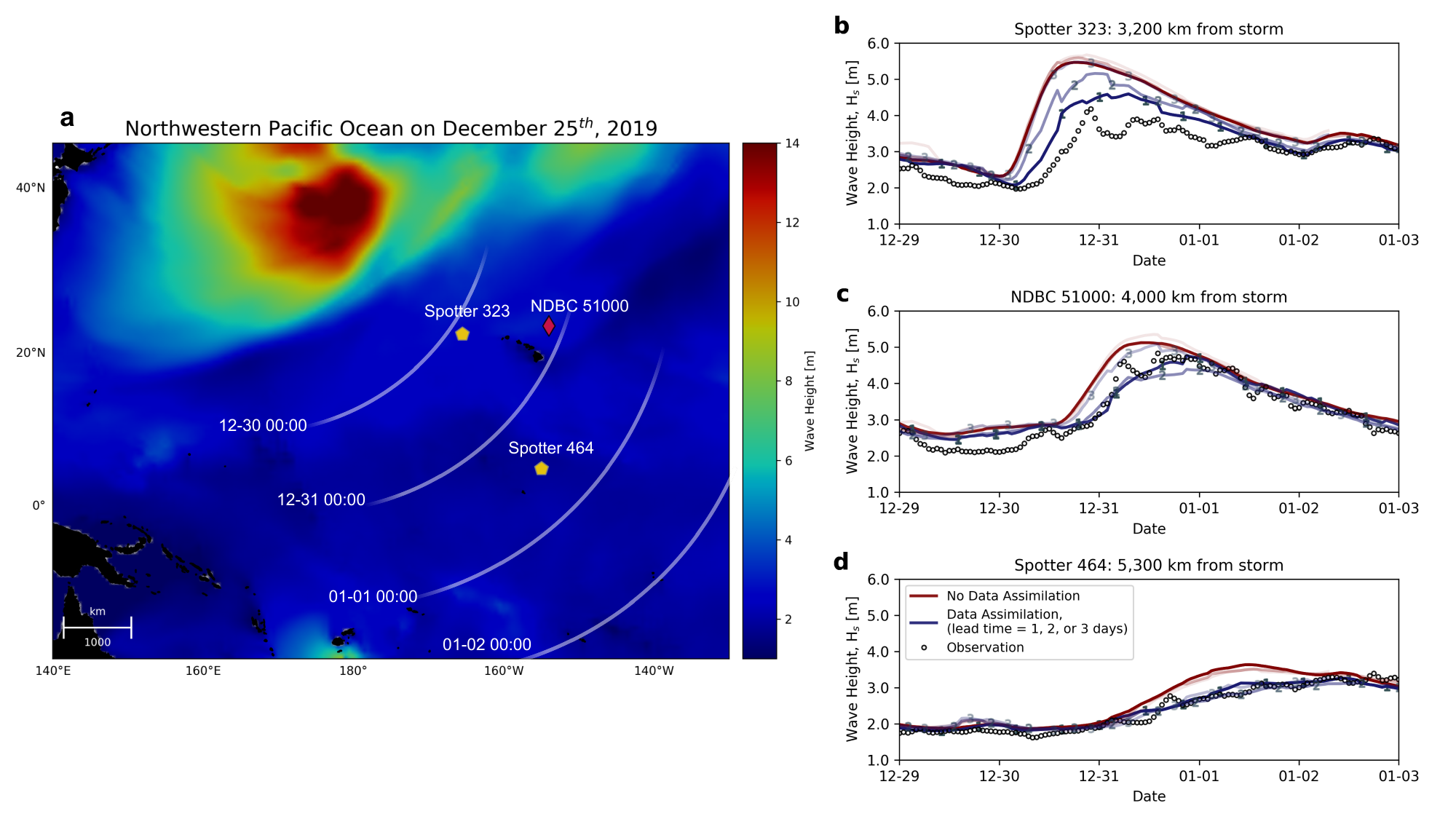}
\caption{(a) Map of the northern Pacific colored by significant waveheight on December 25, 2019. Approximate location of wave swell front (i.e. location of discontinuity in wave period field) is indicated by white curves as it propagates from 12-30-2019 to 1-2-2020. The three buoy locations are indicated on the map, with two Spotters (yellow pentagons) and one National Data Buoy Center buoy (NDBC, red diamond). (b-d) Buoy observations and waveheight predictions with and without data assimilation for lead times of 1, 2, and 3 days. Data assimilation improved forecasts in terms of both swell height and arrival time. Locations further from the storm exhibit additional improvement for longer lead times due to assimilation of more proximate Spotters. }
\label{fig:Decstorm}
\end{figure}

In addition to improvements on the majority of point measurements, data assimilation exhibited increased accuracy for local waveheight time series prediction, where both timing and magnitude is relevant. For example, on December 25$^{th}$, 2019, an energetic storm system developed in the northwestern Pacific about 1,000 km off the coast of  Japan. Surface winds associated with the storm generated waves up to 30 m in height that subsequently radiated outward across the northern Pacific basin (figure \ref{fig:Decstorm}a). The model without data assimilation incorrectly predicted the radiation of the swell across the Pacific, with arrival time errors on the order of 12 hours and waveheight errors of 1-2 meters (figure \ref{fig:Decstorm}b). 

At early times before the storm swell had propagated substantially outward, the model with data assimilation matched the model without data assimilation. However, several Spotters were in the vicinity of the storm, with varying radial distance from the primary wave disturbance. As the most proximate Spotters experienced the storm swell and measured waveheights disparate from model predictions, the model with data assimilation was corrected. This led to unique, and improved, forecasts at Spotter locations further radially outward from the storm, particularly when the forecast lead time allowed for sufficient Spotter data to be taken into account. This was most notable for Spotter-464 (figure \ref{fig:Decstorm}d), where the one-, two-, and three-day forecasts were all consistent and accurate, and approximately 0.75 m different from the model without data assimilation. For Spotter-323, about 2,000 km closer to the storm than Spotter-464, the initial three-day lead time forecast at that location matched the model without data assimilation. However, as the swell approached, the later forecasts with shorter lead times were able to incorporate new data from other Spotters and adjust such that the one-day lead time forecast was approximately 1 m closer in height and 6 hours closer in arrival time to the observed swell than the model without assimilation (figure \ref{fig:Decstorm}b).

\section{Discussion}
In this work, we demonstrate how a large distributed sensor network can immediately provide improvements in our wave forecasting systems, even with a relatively simple assimilation strategy. Refinements in the assimilation scheme will immediately result in further improvements. This could include improved estimates of the background errors and improved distribution of the innovation over the wave spectrum. For example, estimates of the background error could account for spatial and seasonal variability, and potentially use the partitioned spectrum as the state variable. The latter would allow for more elegant distribution of wave energy according to the likelihood of different wave systems. \citep[e.g. ][]{Portilla-Yandun2016OnSystems}. Further, this immediately allows for assimilation of other bulk parameters, such as directions and periods.

The static nature of prescribed error statistics based on long-term averages precludes knowledge of uncertainties for a given forecast. Since short, event-specific uncertainty contributes strongly to forecasting errors, estimating error statistics directly through ensemble forecasts would capture the error-of-the-day statistics more accurately. In addition, such an ensemble-Kalman based system would retain memory of uncertainty as the errors are propagated away from the initial storm, allowing for more effective correction of swell systems away from the source.

In the data assimilation schemes discussed, the wave field, but not the wind field, is updated. However, the wave field is directly influenced by the wind field, and therefore errors in the wave field likely signal similar errors in the wind field. As a consequence, these schemes can improve swell forecasts once the swell is generated (initial value problem), but are less effective in improving wind-sea forecasts. Specifically, if only the initial conditions are modified, but the potentially errant driving forces remain the same, the corrected initial conditions relax quickly back to the inaccurate state. Consequently, improvements in the wave field dissipate with increasing lead time. Refining the assimilation to include correction to the wind field such that it is in balance with the local waves at analysis time may improve performance, but ultimately improved predictions of sea state will require improved wind forecasts. 

\section{Conclusions}
To demonstrate the value of large and pervasive ocean weather sensors we assimilated a large network of drifting buoys in an operational wave forecast system based on the WaveWatchIII model. Our results show that an efficient sequential assimilation strategy (Optimized Interpolation) for bulk wave heights can meaningfully improve wave forecasts. While refinements to the assimilation strategy will likely improve model skill considerably, this first demonstration illustrates the effectiveness of large distributed sensor networks in constraining the now-state and improving model forecast skill.

Comparison of errors across all measurements as well as for a specific storm event in December of 2019 indicated clear improvements resulting from the data assimilation. These overall improvements in both waveheight magnitude and swell arrival time, particularly for more energetic sea states, are invaluable for accurately assessing ocean state with important societal implications (safety at sea, offshore operations, coastal risks). The marked improvement in forecasting from the inclusion of ocean observations shows the considerable value of greater ocean data density, and the potential of low-cost distributed sensor networks. 

\section*{Acknowledgements}
We thank Paul Wittmann who was instrumental in creating the first version of the operational wave modelling framework used in this study, and who unfortunately passed away before publication of this work. We gratefully acknowledge his contributions to this work and beyond, both as a colleague and a friend.

\bibliographystyle{elsarticle-num-names}

\bibliography{references.bib}


\appendix
\section{Correlation Length Scale Sensitivity}
A Gaussian-like 2D correlation function $\bm \rho$ of the form 
\begin{equation*}
    \rho_{m,n} = \exp\left[ -\left(\frac{D(\bm x_m,\bm x_n)}{\lambda}\right)^p\right]
\end{equation*}
has been considered previously  \citep{Lionello1992AssimilationModel,Breivik1994Model,Voorrips1997AssimilationModel,Greenslade2004BackgroundData}, with $1 \le p\le 2$ and $\lambda = O(100$\,km). The power $p$ mostly affects the peakedness, whereas $\lambda$ influences the spatial extent of the Gaussian-like 2D correlation function. Sensitivity to $p$ is low (not shown), and it was therefore set to $p=3/2$ \citep[similar to][]{Voorrips1997AssimilationModel}. 

\begin{figure}[h]
\centering
\includegraphics[width=\linewidth,page=2]{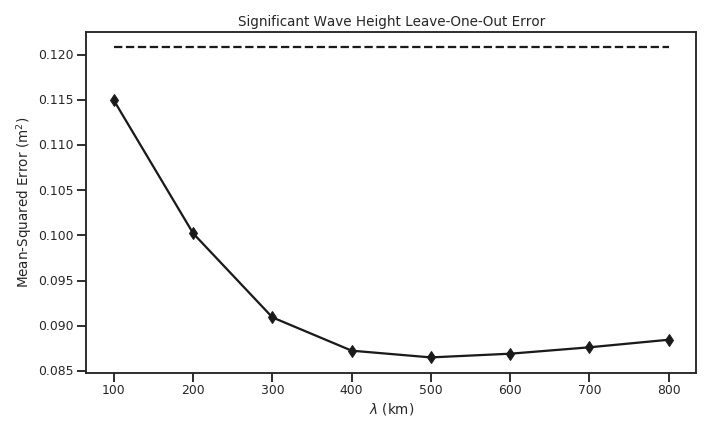}
\caption{Mean-squared error (MSE) for leave-one-out tests for predicted Spotter significant waveheight, as a function of $\lambda$. MSE of the Spotter significant waveheight compared to the forecast is shown for comparison as the dashed line. The minimum predicted Spotter leave-one-out MSE ($0.086$ m$^2$) occurs for $\lambda=500$ km.}
\label{fig:LambdaTests}
\end{figure}

To investigate sensitivity to the length scale parameter $\lambda$, the performance of the optimal interpolation model was evaluated for varying values of $\lambda$ using a leave-one-out evaluation metric. This evaluation calculates the accuracy of the significant waveheight prediction made for each Spotter observer when that observer is excluded from the model's input data. For this evaluation, nowcast results from the unassimilated WW3 model were used to produce OI results at the excluded Spotter locations. This metric was calculated for $\lambda$ values ranging from $100$ km to $800$ km in $100$ km steps. Figure \ref{fig:LambdaTests} shows the mean-squared error for the leave-one-out metric for Spotter observations from July 2019 through December 2019. Based on the evaluation results, the minimum error occurs for $\lambda=500$ km. This agrees with results of \citet{Greenslade2004BackgroundData}, who found $\lambda\approx500$\,km based on satellite altimeter data. However, this evaluation considers data without sequential updating of the base model results (i.e. assimilation from previous time steps do not influence the result). For sequential updates, changes are advected downwave of the region of influence, likely favouring smaller values of $\lambda$. For that reason we set $\lambda$ somewhat conservatively to $300$\,km, a value after the largest decrease in error but before the error function flattens.

\end{document}